# The Use of HepRep in GLAST

J. Perl
*SLAC, Stanford, CA 94025, USA*

R. Giannitrapani, M. Frailis
*Dipartimento di Fisica, Universita degli Studi di Udine - Italy*

HepRep is a generic, hierarchical format for description of graphics representables that can be augmented by physics information and relational properties. It was developed for high energy physics event display applications and is especially suited to client/server or component frameworks. The GLAST experiment, an international effort led by NASA for a gamma-ray telescope to launch in 2006, chose HepRep to provide a flexible, extensible and maintainable framework for their event display without tying their users to any one graphics application. To support HepRep in their GUADI infrastructure, GLAST developed a HepRep filler and builder architecture. The architecture hides the details of XML and CORBA in a set of base and helper classes allowing physics experts to focus on what data they want to represent. GLAST has two GAUDI services: HepRepSvc, which registers HepRep fillers in a global registry and allows the HepRep to be exported to XML, and CorbaSvc, which allows the HepRep to be published through a CORBA interface and which allows the client application to feed commands back to GAUDI (such as start next event, or run some GAUDI algorithm). GLAST's HepRep solution gives users a choice of client applications, WIRED (written in Java) or FRED (written in C++ and Ruby), and leaves them free to move to any future HepRep-compliant event display.

## 1. INTRODUCTION

HepRep is a generic, hierarchical format for description of graphics representables that can be augmented by physics information and relational properties. It was developed for high energy physics event display applications and is especially suited to client/server or component frameworks. The GLAST experiment, an international effort led by NASA for a gamma-ray telescope to launch in 2006, chose HepRep to provide a flexible, extensible and maintainable framework for their event display without tying their users to any one graphics application. This paper describes why GLAST selected HepRep and how they went about implementing a HepRep-based event display in their GAUDI framework.

## 2. GLAST MISSION AND INSTRUMENT

GLAST, the Gamma-ray Large Area Space Telescope [1], measures the direction, energy and arrival time of celestial gamma rays. It consists of a Large Area Telescope (LAT) to measure gamma-rays in the energy range ~20MeV - > 300GeV (there is no telescope now covering this range) and a Gamma-ray Burst Monitor (GBM) to provide correlative observations of transient events in the energy range ~20keV - 20MeV (see Figure 1).

GLAST is scheduled to launch in September 2006 from Florida into an orbit of 550 km, at 28.5 degrees inclination and has a design lifetime of at least 5 years.

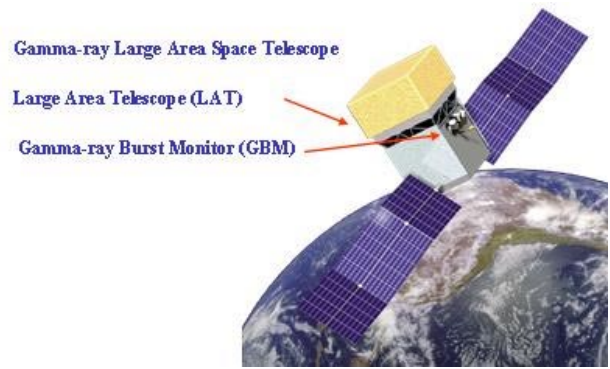

**Figure 1: GLAST Instrument**

### 2.1. Large Area Telescope

The Large Area Telescope (LAT) has an array of 16 identical "Tower" modules, each with a tracker (Si strips) and a calorimeter (CsI with PIN diode readout) and DAQ module (see Figure 2).

The LAT is surrounded by a finely segmented Anti-Coincidence Detector (ACD) (plastic scintillator with photomultiplier tube readout).

GLAST will produce 3GB of data per day. It has 30GB onboard storage, with the data downloaded several times per day. The data eventually makes its way to persistent storage at SLAC.

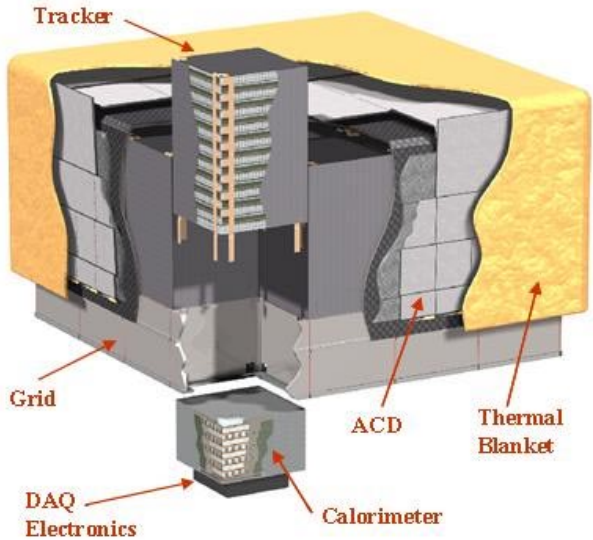

**Figure 2: Large Area Telescope**

## 3. GLAST EVENT DISPLAY REQUIREMENTS

### 3.1. Previous GLAST Event Display

Prior to this work, GLAST already had an integrated Event Display for its offline software (see Figure 3). It had the following features:
- Windows and Linux
- simple, fast wireframe
- 3D and 2D
- tightly integrated into GAUDI [2]
- drives GLAST main offline event loop (source generation, Monte Carlo, reconstruction, etc.)

However, this early version had the following issues:
- very limited mouse interaction
- no ability to pick on objects
- no persistency

For the longer term, GLAST wanted something more flexible, extensible and interactive.

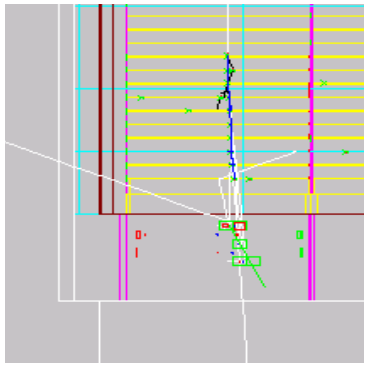

**Figure 3: Previous Event Display**

### 3.2. Requirements for New Event Display

Some key requirements for the new event display were as follows:
- Multiplatform (at least Windows and Linux)
- Easy to install and start
- Fast
- 3D and 2D
- Modern GUI
- Easy navigation and browsing of the event (with incremental download)
- Pick on objects to inquire about them
- Pick on objects to interact with physics algorithms in the GAUDI framework
- Ability to drive the GLAST main offline event loop (source generation, Monte Carlo, reconstruction, etc.)
- Extensible/configurable by the user such that requirements dictated by physics can be easily added directly by the physics experts (should not require graphics experts).
- New features related to what is represented (for example changing the trajectory color to code for charge, or energy, or any other attribute) should not require significant re-coding.

### 3.3. Software Life-Cycle Issues

The correct event display design must assume that the life-cycle of the event display may be different from the life-cycle of the infrastructure software. Therefore, the display should not be too tightly coupled with GLAST's actual choices of:
- framework
- physics algorithms
- event structure
- persistency mechanisms
- Monte Carlo
- etc.

These requirements led GLAST towards an event display paradigm rather than a specific event display application.

### 3.4. The Client-Server Paradigm

An answer to GLAST's requirements is the client-server paradigm. The server deals with physics, interaction with reconstruction algorithms, event store, etc. The client deals only with graphics representations (that may be augmented with additional information that has meaning for the experiment).

The client-server model does not necessarily imply remote operation. Client and server may be on the same machine or may be on different machines. The client-server separation is in any case a useful construct to cleanly delineate the two parts of the event display solution.

The client and server communicate with an interface. This interface should be simple and extensible and should accommodate all the requirements listed above. HepRep is such an interface.

## 4. INTRODUCTION TO HEPREP

HepRep[3,4] is a generic interface for component or client-server event displays. It provides for the correct distribution of computing work between the two parts of the system and effectively addresses the many important maintenance issues involved in such a system.

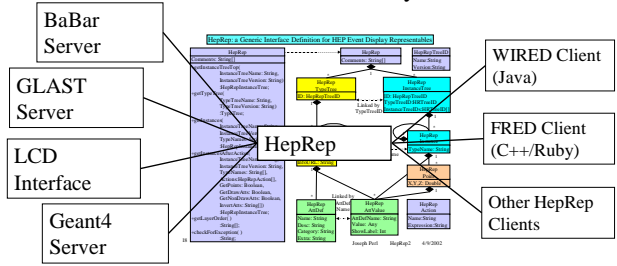

**Figure 4: HepRep Breaks the Dependency**

The HepRep interface breaks the dependency between any particular experiment's event display server and any particular event display client. The HepRep format is independent of any one particular language or protocol. It can be used from C++ or Java and can be shipped as CORBA, RMI, XML, C++, Java or JNI for consumption by WIRED[5,6], FRED[7] or any other HepRep-enabled event display client (see Figure 4).

The rest of this section constitutes a short introduction to HepRep. For a more complete discussion, see http://heprep.freehep.org.

### 4.1. Representables

A naive implementation of a client-server event display would have one simply ship a reference to the physics object. The client could then use a remote interface to make any necessary calls on that physics object. However, such an approach would have an excessive call overhead, as one asks track by track and hit by hit for various coordinate points. Also, such an approach doesn't achieve good separation of client-server functionality.

The design decision behind HepRep is to serve "representables", not physics objects. A representable is the essential spatial information of a physics object (track, calorimeter hit, etc.) and can be augmented by that object's physics attributes (momentum, energy, etc.).

Serving Representables keeps the detailed reconstruction code, swimmers and detector models on the server side where they belong. Spatial information is assembled and shipped in an efficient manner, avoiding the overhead of too many individual method calls. Rendering decisions are deferred, as much as possible, to the client.

### 4.2. Example Representable: Track

A precise fitted track could be served as a set of swim step points, each augmented by helix parameters and descriptive information (track number, particle id, etc.)(see Figure 5). Only in the client is the final decision made whether to represent this representable as
- a dotted line,
- or as set of individual swim step momentum vectors,
- or as a set of helix segments.

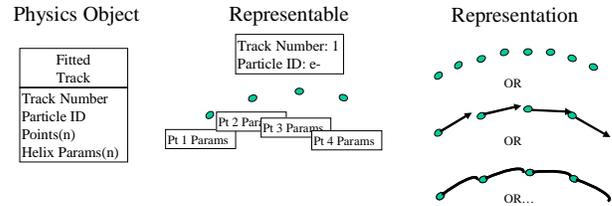

**Figure 5: Example Representable - Track**

### 4.3. Example Representable: Calorimeter Crystal

A calorimeter crystal is served as the corner points of a prism shape, augmented by an energy value and other information. Only in the client is the final decision made whether to represent this representable as
- a realistically sized crystal,
- or as an energy tower with a base at the real crystal position but a length sized by energy,
- or as some other shape.

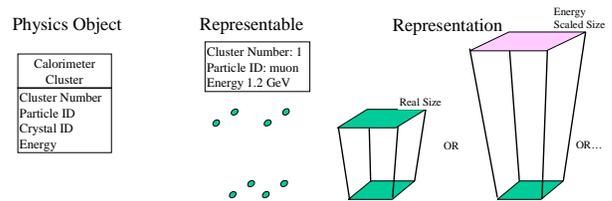

**Figure 6: Example Representable - Cal Crystal**

### 4.4. HepRep Object Tree

HepRep representables are arranged in a tree. An example HepRep object tree is shown in Figure 7. The tree has two main parts: the Type Tree, which describes characteristics common to all instances of a given type (all tracks, all calorimeter clusters, etc.) and the Instance Tree, which describes the specific instances of a given type (one instance for each track, etc.).

Another key element of HepRep is its very flexible scheme for incremental downloads. A client can ask to:
- include or exclude Attributes
- only get Instances of a given Type
- only get Instances that have given Attributes
- get Instances according to other options

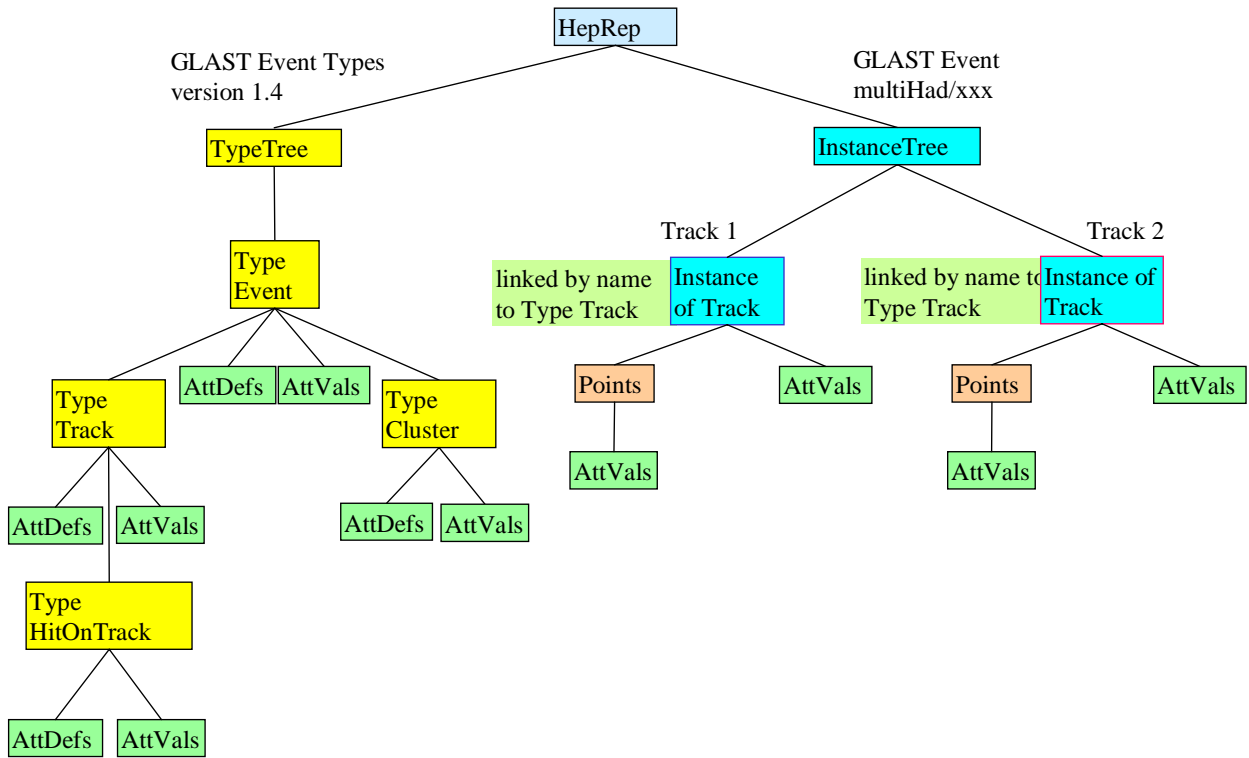

**Figure 7: HepRep Object Tree**

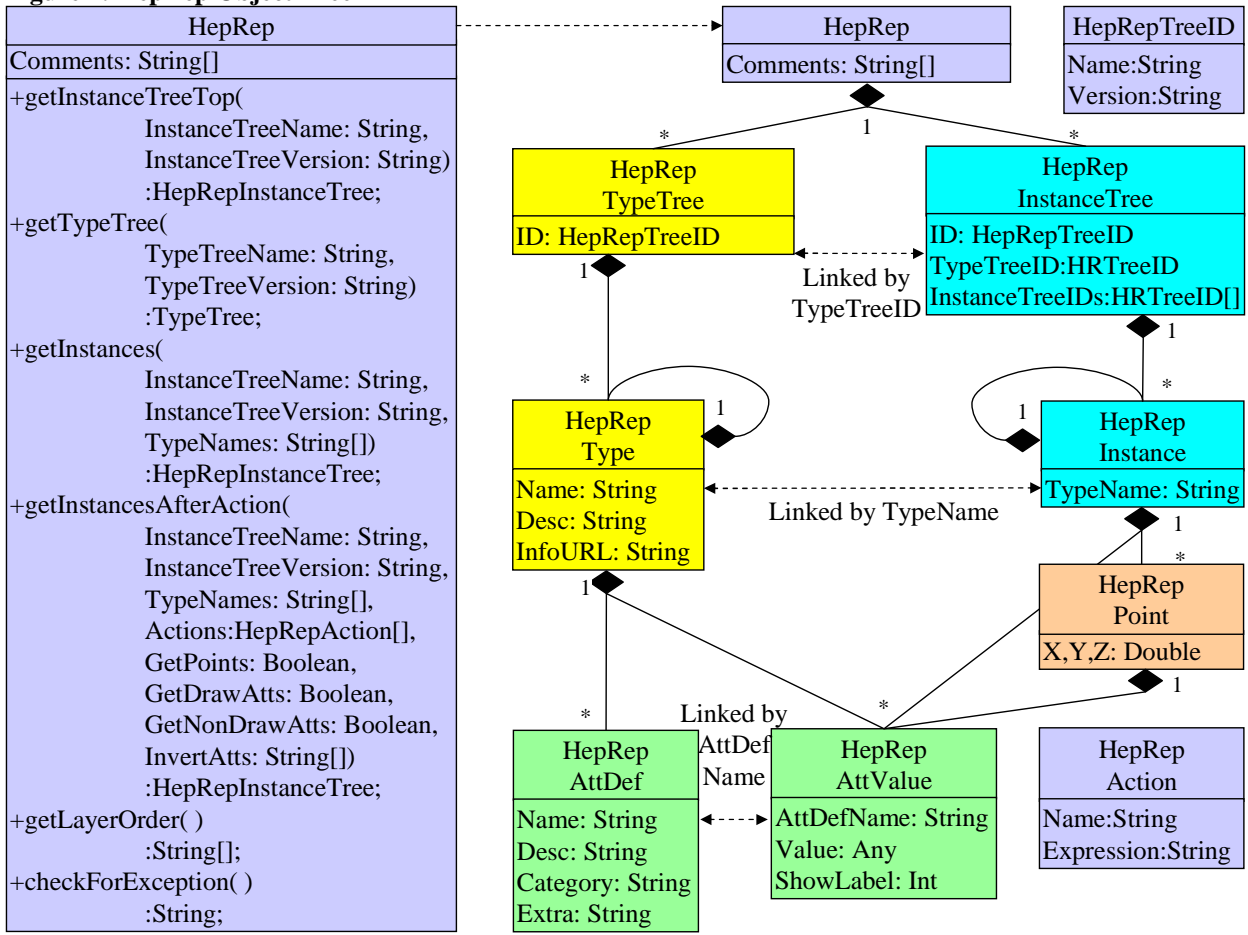

**Figure 8: HepRep UML Diagram**

**THLT009**

## 4.5. HepRep Complete UML Diagram

The HepRep object contains two main trees, the TypeTree and the InstanceTree.

The TypeTree has zero or more Types (such as track, cluster, etc.). Types can in turn have subTypes (such as the Hits on a Track). Each Type can also have attributes, which may be predefined (drawing attributes such as color, line width, etc.) or may be defined for that specific type (such as the momentum attribute for a Track).

The InstanceTree has zero or more Instances (individual tracks, clusters, etc). Instances can in turn have subInstances (such as the specific Hits on a specific Track). Each Instance can have zero or more points (the space points used to draw a point, line, polygon or other graphics primitive). Each Instance can also have attributes (which override the defaults set in the relevant Type for all Instances of the given Type).

The HepRep's four principle methods (getInstanceTreeTop, getTypeTree, getInstances and getInstancesAfterAction) provide the flexibility to let the client application discover what Types are available but only download those Instances and Attributes that are of current interest. For a more complete discussion, see http://heprep.freehep.org.

## 4.6. HepRep Attributes

The full power of HepRep comes from its simple yet very flexible system of Attributes. Some of the drawing attributes are predefined, but any other necessary attributes can be defined as needed for a specific HepRep Type (such as the Momentum attribute of a Track).

Any number of Attributes can be hung from a Type, Instance or Point (see Figure 9).

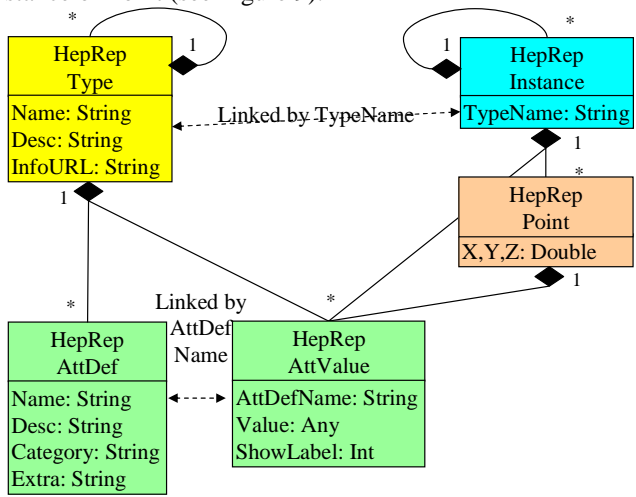

**Figure 9: HepRep Attributes**

There are four Categories of Attributes:
- Draw Attributes (such as thickness, color and what shape to draw from the points) can be modified in the client through a draw attribute editor
- Physics Attributes (such as track momentum or hit error) can be used for visibility cuts (client side or server side)
- PickAction Attributes define special things to do when the user picks on the Representable (such as remove hit and refit track)
- Association Attributes define loose associations between Representables (such as track cluster matching)

## 5. THE GLAST WAY TO HEPREP

GLAST uses GAUDI, an object oriented (C++) framework that has a strong separation between data and the algorithms on that data. Data are stored in the Transient Data Store (TDS) and/or the Permanent Data Store. Algorithms can act on the TDS, filling it or retrieving things from it. Services provide common functionalities on algorithms.

GAUDI has its own event loop which can be customized at runtime through initialization files (jobOptions files). GLAST needed to develop its client-server HepRep framework inside GAUDI in such a way that one could drive the event loop from the external graphics client
- The server component lives in the GAUDI world, having access to all Algorithms, Services and the TDS. It has full knowledge of the physics contents of each event.
- The client component lives outside GAUDI. It implements the graphical user interface and has access to HepRep attributes, but does not have direct access to tools within GAUDI.
- The HepRep interface brings information from the server to the client and brings commands from the client back to the server. HepRep can be implemented in various ways. GLAST currently has implementations in XML (persistent) and CORBA (live).

### 5.1. HepRepSvc and CorbaSvc

The GLAST event display solution involves two GAUDI services: HepRepSvc and CorbaSvc.

At the end of each event, the HepRepSvc produces a HepRep representation of the event from the TDS. This representation can be published either as a persistent XML file or as a CORBA object. The XML file may contain Instances of more or fewer HepRepTypes depending on user job options. To minimize unnecessary memory costs, the design takes care that the entire HepRep is never actually held in C++ memory but is instead streamed directly out to XML or CORBA.

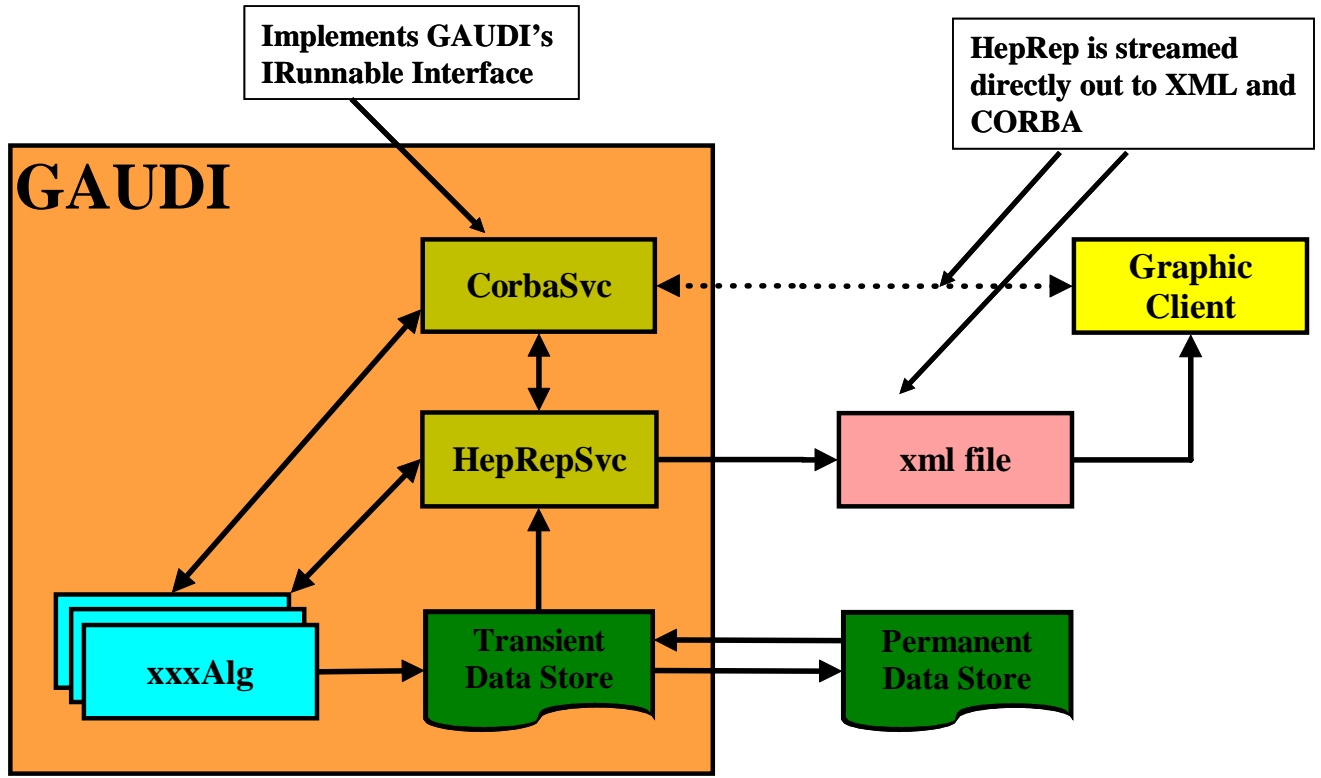

Figure 10: HepRepSvc and CorbaSvc

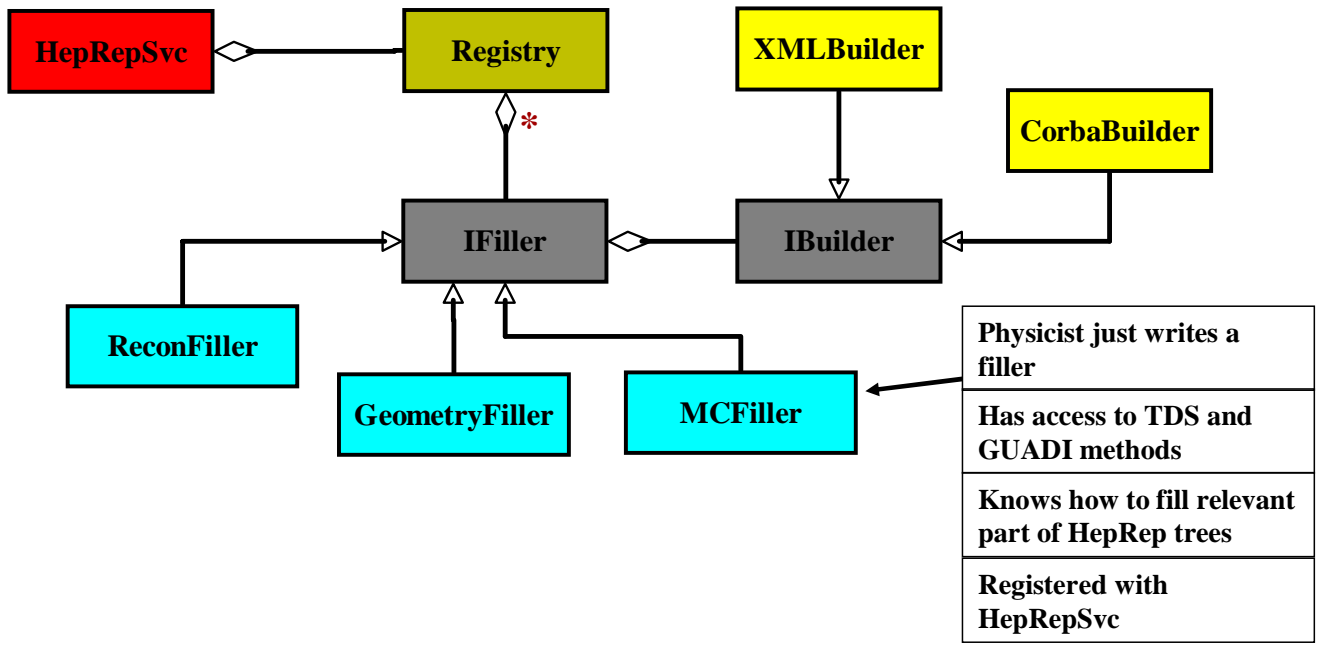

Figure 11: HepRep Filler and Builder Architecture

The CorbaSvc implements a GAUDI IRunnable interface and so can drive the event loop. It publishes a CORBA HepRep object that can be used by the graphics client to retrieve the event. The client can ask for Instances of more or fewer HepRep Types, exploiting the full flexibility of the HepRep download methods.

Since the CorbaSvc lives inside GAUDI, it is possible (at least in principle) to call all of the collaboration's algorithms on the TDS event. This framework allows complete interaction of the graphics client with the physics software. Since it is an IRunnable, the CorbaSvc can stop the event loop and wait for remote method invocations from the external client (see Figure 10).

### 5.2. The HepRep Filler and Builder Architecture.

The remaining issue is exactly how the HepRep is built out of the TDS. The goal is to make it easy for physics experts to add whatever they want into the HepRep hierarchy without requiring the assistance of event display experts and without requiring the individual physics experts to coordinate with each other.

The solution, borrowed in part from the BaBar collaboration, is the HepRep filler mechanism. A base HepRep filler class (IFiller) provides a variety of convenience methods to create and fill parts of the HepRep structure. It has methods to create HepRep Types and Instances, linking them into the proper hierarchy; it has methods to convert various types of data to HepRep attributes.

For each kind of physics data to be represented (each type of Representable), the physics expert just creates a concrete instance of this IFiller. He has access to the TDS and to all relevant GAUDI methods and to tools to retrieve information from the TDS. He prepares the relevant information from the event and then, with a few simple calls to the base class, fills that information into his piece of the HepRep TypeTree and InstanceTree.

Whereas BaBar's HepRep filler base class makes direct calls to a CORBA HepRep layer, the GLAST HepRep filler takes the abstraction one step further. The GLAST filler uses an abstract HepRep Builder (HepRep factory) for which the concrete HepRep implementation can be either XML or CORBA. Thus with no extra work on the part of the physics expert, output can be persistent (XML) or live (CORBA) (see Figure 11).

Each filler is listed in a register (held by HepRepSvc) with the filler associated to one or more HepRep Types. At each event, depending on what Types the user has asked to see, the relevant fillers are called to fill the desired parts of the HepRep.

### 5.3. Value of the Filler and Builder Architecture.

The Filler and Builder architecture provides good separation between the physics experts and the graphics experts. In practice, the physicist just looks at a few example fillers and then makes a new one based on those examples. The physicist does not need to learn anything else about the event display server or client (no need to learn CORBA, no need to learn XML, no need to learn Java or Ruby)

Different physicists can work on different fillers. The architecture provides good separation between GLAST subsystems, either physical (for example Anti-Coicidence Detector from Tracker) or conceptual (for example reconstruction from Monte Carlo).

The architecture provides good abstraction from the concrete implementation of the HepRep. One filler is used for all the possible HepRep formats (CORBA, XML, etc.).

The architecture is very flexible. New fillers can be added at any time. A similar filler mechanism has been used by BaBar to good effect.

### 5.4. Relationship between GLAST's Use of HepRep and that of Other HepRep Users

HepRep is currently being used by BaBar[8], GLAST, LCD[9] and Geant4[10], and there are two HepRep client applications, WIRED and FRED. While all four groups are now using WIRED, and two can use FRED (the other two need to migrate to HepRep version 2 before they can work with FRED), the four groups use a variety of HepRep and legacy implementations (see Figure 12):

- BaBar has a HepRep1 CORBA server..
- LCD passes WIRED java objects using a legacy data format (pre-HepRep).
- Geant4 has abstract HepRep1 and HepRep2 factories with implementations to XML and Java.
- GLAST has a different abstract HepRep2 factory with implementations to XML and CORBA.

In the near future, all data sources will speak HepRep2 to an abstract HepRep factory (from the FreeHEP Software Library [11]). By instantiation of one or another concrete implementation of HepRep, a C++ program will be able to change from creating HepRep in C++ memory to creating HepRep as an XML streamer (a pure C++ solution with no external library dependencies and no creation of the HepRep in memory) to creating HepRep as a CORBA streamer (depends on CORBA libraries) or creating HepRep as Java (via Java Native Interface) (see Figure 13).

**THLT009**

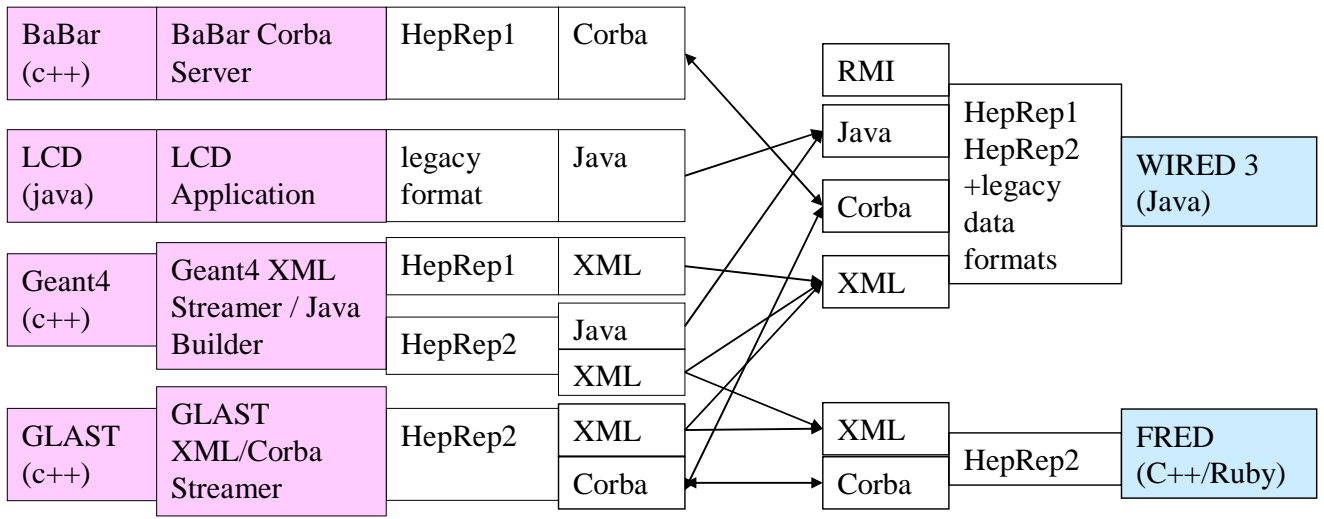

**Figure 12: HepRep Current Use Architecture**

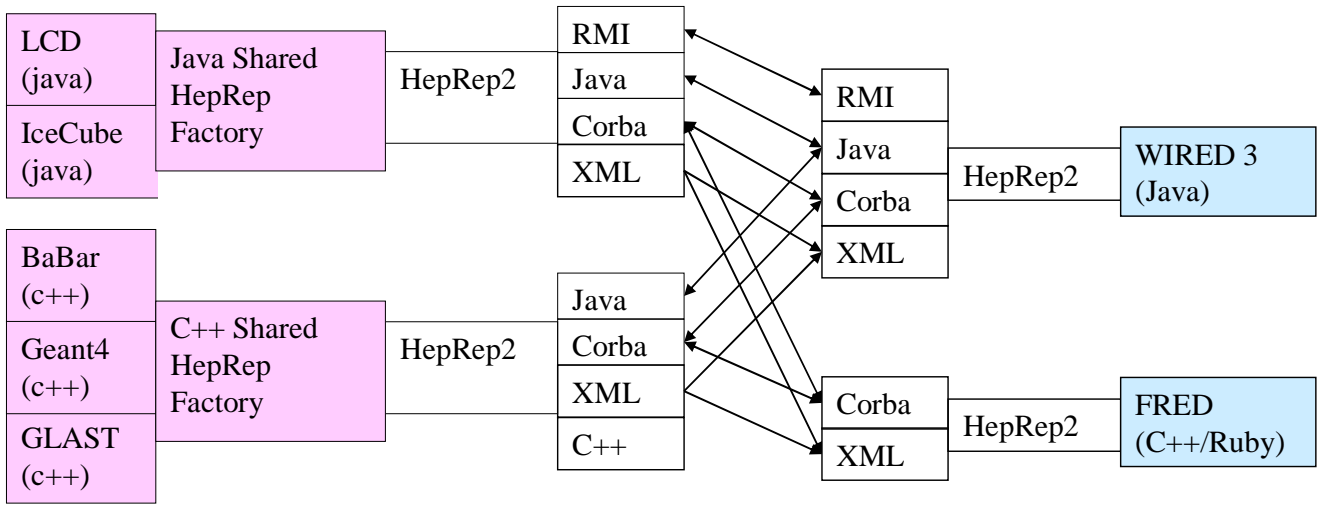

**Figure 13: HepRep Near-Term Future Architecture**

## 6. CONCLUSIONS AND OUTLOOK

GLAST wanted a flexible, extensible and maintainable framework for event displays without committing to any one graphics application. GLAST has accomplished that with a HepRep based client-server framework integrated into their GAUDI application. The implementation uses a filler and builder mechanism to abstract event description from event representation (allowing the physics expert to extend the display with no special knowledge of CORBA, XML, Java or Ruby).

GLAST's event display implementation has given their users a choice of client application:

- WIRED, based on Java (see Figure 14)
- FRED, based on C++ and Ruby (see Figure 15)

Because GLAST has chosen a component architecture with a well defined, language neutral, interface, HepRep, the GLAST user is free to choose any event display application that implements HepRep.

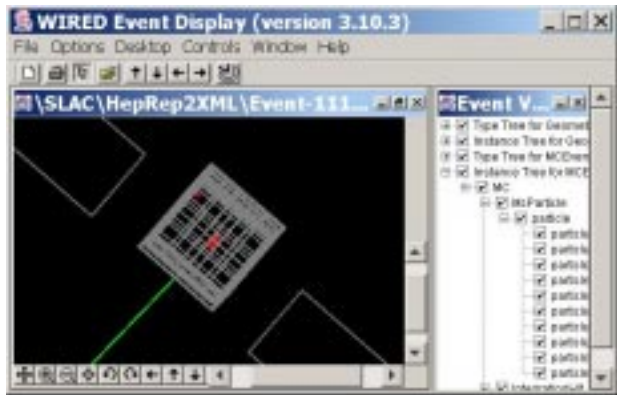

**Figure 14: GLAST in the WIRED Event Display**

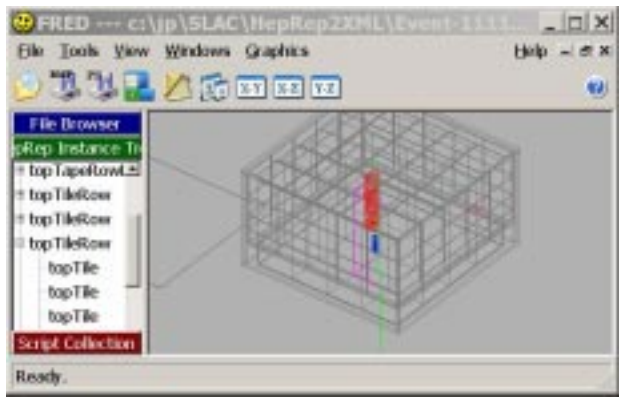

**Figure 15: GLAST in the FRED Event Display**

## Acknowledgments

Work supported by Department of Energy contract DE-AC03-76SF00515.